\documentclass[]{aa}
\usepackage{graphicx}
\usepackage{txfonts}
\begin{document}
   \title{HST/FOS Time-resolved spectral mapping of IP~Pegasi at the end of an outburst}

%   \subtitle{I. Overviewing the $\kappa$-mechanism}

   \author{R. K. Saito,
          \inst{1}\fnmsep
	  \thanks{\textit{Correspondence to} saito@astro.ufsc.br}
	  R. Baptista\inst{1}
          \and
          K. Horne\inst{2}
          }

%   \offprints{G. Wuchterl}

   \institute{Departamento de F\'{\i}sica, Universidade Federal de Santa Catarina, 
              Trindade, 88040-900, Florian\'opolis, SC, Brazil\
%              \email{saito@astro.ufsc.br}
         \and
             School of Physics and Astronomy, University of St. Andrews, KY16 9SS, 
             Scotland, UK\
%             \email{c.ptolemy@hipparch.uheaven.space}
%             \thanks{The university of heaven temporarily does not
%                     accept e-mails}
             }

   \date{Received September 15, 1996; accepted March 16, 1997}

   \abstract{

We report an eclipse mapping analysis of time-resolved ultraviolet
spectroscopy covering three eclipses of the dwarf nova IP~Pegasi on
the late decline of the 1993 May outburst. The eclipse maps of the
first run show evidence of one spiral arm, suggesting that spiral
structures may still be present in the accretion disc 9 days after the
onset of the outburst.  In the spatially resolved spectra the most
prominent lines appear in emission at any radius, being stronger in
the inner disc regions. The spectrum of the gas stream is clearly
distinct from the disc spectrum in the intermediate and outer disc
regions, suggesting the occurrence of gas stream overflow.  The full
width half maximum of C\,IV is approximately constant with radius, in
contrast to the expected $v\propto{R^{-1/2}}$ law for a gas in
Keplerian orbits. This line probably originates in a vertically
extended region (chromosphere + disc wind).  The uneclipsed component
contributes $\sim{4}$ \% of the flux in C\,IV in the first run, and
becomes negligible in the remaining runs.  We fit stellar atmosphere
models to the spatially resolved spectra.  The radial run of the disc
color temperature for the three runs is flatter than the expected
$T\propto{R^{-3/4}}$ law for steady-state optically thick discs
models, with $T\simeq{20000}$ K in the inner regions and
$T\simeq{9000}$ K in the outer disc regions.  The solid angles that
result from the fits are smaller than expected from the parameters of
the system.  The radial run of the solid angle suggests that the disc
is flared in outburst, and decreases in thickness toward the end of
the outburst.

    }

   \maketitle
%
%________________________________________________________________

\section{Introduction}

Accretion discs are cosmic devices that allow matter to efficiently
accrete over a compact source by removing its angular momentum via
viscous stresses while transforming gravitational potential energy
into head and, thereafter, radiation (Frank, King \& Raine
1992). Dwarf novae are excellent environments for advances in the
knowledge of accretion physics. In these close binaries mass is fed to
a white dwarf (the primary) by a Roche lobe filling companion star
(the secondary) via an accretion disc which usually dominates the
ultraviolet and optical light of the system (Warner 1995).

The observation of accretion discs in dwarf novae is well beyond the
capabilities of current direct-imaging techniques as it requires
resolving structures on angular scales of micro-arcseconds. However,
the binary nature of these systems allows the application of powerful
indirect-imaging techniques such as eclipse mapping (Horne 1985) and
Doppler tomography (Marsh \& Horne 1988) to probe the dynamics,
structure and the time evolution of their accretion discs
(e.g. Baptista 2001).

IP~Pegasi is an intensively studied deeply eclipsing dwarf nova ($\rm
P_{orb}= 3.8$~hr) which shows 1-2 weeks-long, $\simeq 2$~mag outbursts
every 60-120 days. In the last decade, several studies, many of them
based on indirect-imaging techniques, revealed interesting results
about the structure and behavior of the IP~Peg accretion disc.
Bobinger et al. (1997) performed optical eclipse mapping of IP~Peg on
decline from an outburst to find that the temperature distribution was
much flatter than expected for a steady-state disc, with brightness
temperatures in the range T $\simeq9000-5000$ K for the inner and
outer disc regions, respectively.

The discovery of spiral arms in the accretion disc of IP~Pegasi during
outbursts with Doppler tomography (Steeghs, Harlaftis \& Horne 1997)
confirmed the results of hydrodynamical simulations (Stehle 1999). The
spiral arms appear because of shocks caused by the tidal effect of the
secondary star on the outer parts of the extended accretion disc
during outbursts. Steeghs et al. (1997) and Morales-Rueda, Marsh \&
Billington (2000) confirmed that the spiral structure was a stable
feature of IP~Peg during outbursts, and that it was still present 5-7
days after the onset of the outburst. Baptista, Harlaftis \& Steeghs
(2000) and Baptista, Haswell \& Thomas (2002) [hereafter BHS and BHT,
respectively] confirmed the presence and recovered the spatial
location of spiral arms on the outburst decline from optical eclipse
mapping.

Here we report the results of a time-resolved ultraviolet (UV)
spectral mapping experiment of IP~Pegasi 8-11 days after the onset of
the May 1993 outburst, when the star's optical brightness was almost
back to its quiescent level. The analysis is based on observations
secured with the Faint Object Spectrograph onboard the Hubble Space
Telescope.  The observations and the data reduction are described in
Section 2. The details of the data analysis are given in Section 3.
In Section 4 we investigate the disc spatial structure in the lines
and in the continuum and we present spatially resolved spectra of the
disc, gas stream and the spectrum of the uneclipsed light. We fit
stellar atmosphere models to the disc spectra in order to derive the
radial run of the temperature and solid angle of each disc surface
element. The results are summarized in Section 5. In the Appendix we
address the reliability of eclipse mapping reconstructions with light
curves of incomplete phase coverage.

%__________________________________________________________________

\section{Observations}

\begin{figure}
\includegraphics[bb=1.5cm 2cm 20cm 18cm,angle=-90,scale=.4]{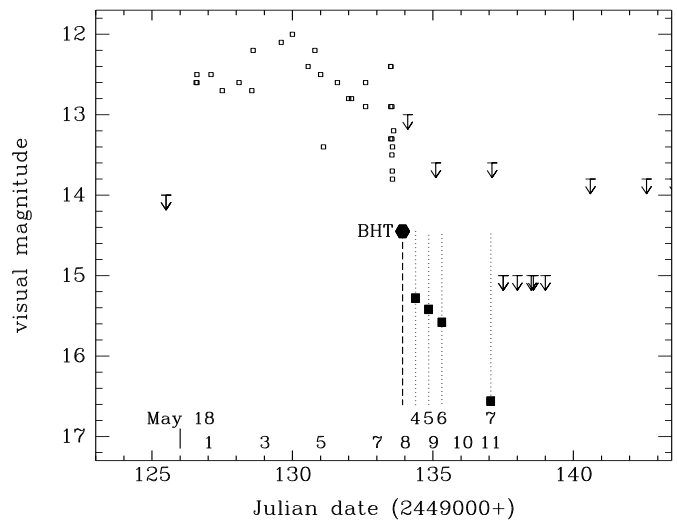}
 \caption{ Visual light curve of IP~Peg for the epoch 1993 May-June,
   constructed from observations made by the AAVSO (open
   squares). Arrows indicate upper limits on the visual magnitude.
   The synthetic V-band out-of-eclipse magnitude from the data of BHT
   is shown as a filled hexagon. A vertical dashed line marks the
   epoch of their observations. Vertical dotted lines mark the epoch
   of our observations. AB$_{79}$ magnitudes in the range $1100-2500$
   \AA\ are shown as filled squares for illustration purposes [labeled
   runs 4, 5, 6 and 7 (IP+)].  }
 \label{erupcao}
\end{figure}

IP Peg went in outburst on 1993 May 18. The Faint Object Spectrograph
(FOS) onboard the Hubble Space Telescope was used to secure
time-resolved spectroscopy covering 4 eclipses in 1993 May 27-30,
during the late decline from outburst. Fig.\,\ref{erupcao} shows the
historical optical light curve of IP~Peg around the 1993 May
outburst. The times of the HST runs are indicated by vertical dotted
lines. The observations of BHT were performed roughly one day before
the start of the HST runs and are also indicated in
Fig.\,\ref{erupcao}. Therefore, the HST observations allow us the
opportunity to track the changes in the disc of IP~Peg in the days
following the observations reported by BHT.

The HST observations, detailed in Table 1, were performed in the
'rapid readout' mode at a time resolution of 5.5\,$s$.  To increase
the nominal spectral resolution of the FOS spectra, during the
acquisition the spectrum was shifted electronically by a quarter of a
diode along the 516--diode Digicon array ('1/4 sub-stepping')
resulting in a 2064-pixel spectrum with a net exposure time of
1.185\,s pixel$^{-1}$. The runs were performed with the $4.3\times4.3$
arcsec$^{2}$ aperture and the G160L grating ($\lambda 1100-2500$~\AA,
$\Delta\lambda= 1.7$ \AA\ pixel$^{-1}$).  Because of the relatively
large slit aperture, these spectra are contaminated by geo-coronal
Ly$\alpha$ emission. The choice of the slit aperture was dictated by
the large point-spread function of HST at that epoch (these are
pre-COSTAR observations). The runs cover the eclipse cycles 22246
(IP4), 22249 (IP5), 22252 (IP6) and 22263 (IP7), according to the
linear ephemeris of Wolf et al. (1993).

\begin{figure*}
\includegraphics[bb=2cm -2cm 20cm 24cm,angle=-90,scale=0.6]{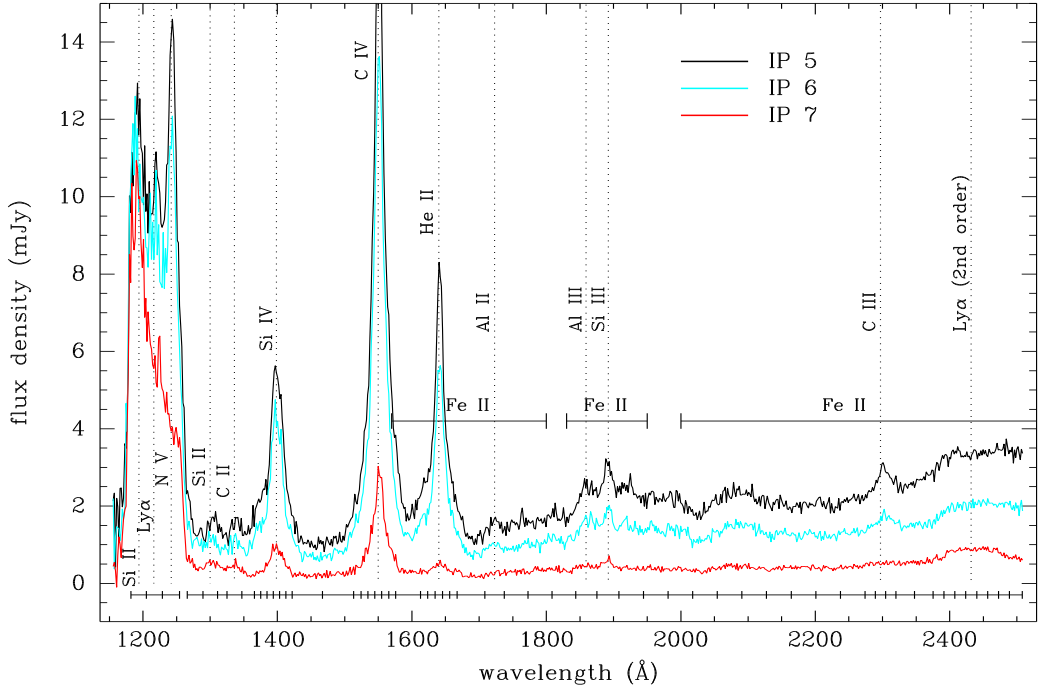}
 \caption{Average out-of-eclipse spectra of IP~Peg during the decline
of the 1993 May outburst (phase range +0.10 to +0.14 cycle). The
observations correspond to the eclipse cycles 22249 (IP5), 22252 (IP6)
and 22263 (IP7), according to the ephemeris of Wolf et
al. (1993). Vertical dotted lines mark the emission lines. Horizontal
lines mark the Fe\,II absorption bands.  Horizontal ticks indicate the 58
narrows passbands used to extract light curves.}
 \label{spec_med}
\end{figure*}

\begin{table}[h]
\begin{center}
\caption{Journal of the observations.}
\begin{tabular}{c c c c c}  
  % after \\: \hline or \cline{col1-col2} \cline{col3-col4} ... %
      & Date & UT & & \\
  Run & (1993 May) & (Start) & Eclipse & Phase Range \\ \hline
  IP4 & 27 & 09:05 & 22246 & $+0.050, +0.212$ \\
  IP5 & 27 & 20:20 & 22249 & $+0.011, +0.174$ \\
  IP6 & 28 & 07:34 & 22252 & $-0.028, +0.136$ \\
  IP7 & 30 & 01:20 & 22263 & $-0.028, +0.136$ \\
\end{tabular}
\end{center}
\end{table}

Due to problems with the HST scheduling, the first run (IP4) missed
the eclipse. Since that data set contains only the late egress of the
light curve it will not be included in the following
analysis. Unfortunately, because of the same scheduling problem, the
other runs are also miscentered and do not cover the ingress of the
eclipse (see Fig.\,\ref{curvas}).

The observations were reduced with a procedure similar to the standard
STSDAS pipeline and included flat-field and geomagnetically induced
motion ('GIMP') corrections, background and scattered light
substraction, wavelength and absolute flux calibrations.

The average out-of-eclipse UV spectra of runs IP5, IP6 and IP7 are
shown in Fig.\,\ref{spec_med}. They show emission lines of Si\,II
$\lambda 1194$, Ly$\alpha$ $\lambda 1216$ (mostly geo-coronal), N\,V
$\lambda 1240,1243$, Si\,II $\lambda 1300$, C\,II $\lambda 1336$,
Si\,IV $\lambda 1394,1403$, C\,IV $\lambda 1549,1551$, He\,II $\lambda
1640$, Al\,II $\lambda 1723$, Al\,III $\lambda 1859$, Si\,III $\lambda
1892$, C\,III $\lambda 2297$, as well as broad absorption bands
possibly due to Fe\,II (e.g. see Fig.\,\ref{spec_disc}).  The
strong and broad emission feature centered at $\simeq 1210$\,\AA\ is
due to geo-coronal $\rm Ly\alpha$ emission filling all the relatively
wide slit aperture. Narrow Si\,II, $\rm Ly\alpha$ and N\,V line
emission from IP Peg are seen on top of the broad geo-coronal
component. The position of the second-order $\rm Ly\alpha$ emission is
also illustrated in Fig.\,\ref{spec_med}. The broad hump seen in the
spectrum at this wavelength (particularly in IP7) suggests a
non-negligible contamination by geo-coronal $\rm Ly\alpha$ at these
wavelengths too.

The continuum emission decreases by a factor of 3 from run IP5 to run
IP7. The decrease in line strength is more pronounced, with most of
the lines roughly disappearing in the continuum at IP7.

%_____________________________________________________________

\section{Data analysis} 

\subsection{Light-curve construction}

The spectra were divided into a set of 58 narrow passbands, with 23
continuum passbands 16-50 \AA\ wide and 35 passbands for 13 lines
(Fig.\,\ref{spec_med}). The strongest emission lines were separated in
velocity-resolved bins 2000~km/s wide, with a bin centered in the rest
wavelength ($v=0$\,km/s). For the fainter lines we defined a single
bin of $5000-6000$\,km/s centered in the rest wavelength. The systemic
velocity of IP Peg ($\gamma\simeq 30$\,km/s) [Martin et al. 1989] is
both rather uncertain and much smaller than the width of the passbands
and was neglected. The S/N of the spectra decreases towards shorter
wavelengths, which prevented us from extracting a usable light curve
blueward of the $\lambda 1194$ line.

For those passbands including emission lines the light curve comprises
the total flux at the corresponding bin, with no substraction of a
possible continuum contribution.

Light curves were extracted for each passband by computing the average
flux on the corresponding wavelength range and phase-folding the
results according to the linear ephemeris of Wolf et al. (1993),

\begin{equation}
T_{mid}(HJD)=2445615.4156+0.15820616E,
\end{equation} 
where $T_{mid}$ gives the superior conjunction of the white dwarf. The
error bars are taken as the standard derivation with respect to the
average flux at each phase bin.

IP~Peg shows long-term ($\simeq$ 5yr) cyclical orbital period changes,
with departures of the observed mid-eclipse timings of up to 2 \rm{min}
with respect to the ephemeris of equation (1). Fortunately, the HST
outburst observations were bracketed by two observations of IP Peg in
quiescence in which the egress time of the white dwarf can be easily
measured (Baptista et al. 1994). From the contemporary quiescent data
we inferred a white dwarf mid-eclipse phase of
$\phi_{o}=-0.0082$\,cycle for this epoch (adopting an eclipse width of
$\Delta\phi=0.0863$\,cycle, Wood \& Crawford 1986) and corrected the
data accordingly to make the centre of the white dwarf eclipse
coincident with phase zero. 

\begin{figure}
  \centering
    \includegraphics[bb=4cm 6cm 18cm 20cm,angle=-90,scale=.43]{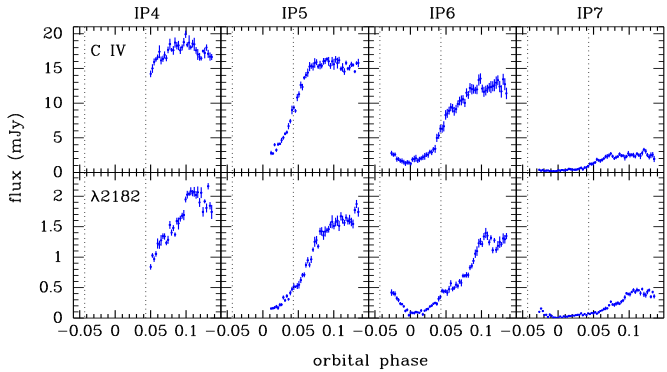}
    \caption{Light curves for the C\,IV emission line centre and for a
             selected continuum passband. Vertical dotted lines mark
             the ingress/egress phases of the white dwarf.}
        \label{curvas}
  \end{figure}

Light curves at a selected continuum passband and for the C\,IV line
centre passband are shown in Fig.\,\ref{curvas}. The incomplete
eclipse phase coverage of the runs is clear. Light curves for the IP4
run are also shown. The eclipse in the IP4 continuum is perceptibly
wider than in the corresponding C\,IV light curve and the egress
shoulder in the continuum eclipse shape indicates the reappearance of
an asymmetric bright structure, possibly the ``red'' spiral arm (see
Section 4.1).

\subsection{Eclipse mapping}

Maximum-entropy eclipse mapping techniques (Horne 1985, Baptista \&
Steiner 1993) were used to solve for a map of the disc brightness
distribution and for the flux of an additional uneclipsed component in
each band. The reader is referred to Baptista (2001) for a recent
review on the eclipse mapping method.

As our eclipse map we adopted a flat grid of $51\times51$ pixels
centered on the primary star with side $2\,R_{L1}$, where $R_{L1}$ is
the distance from the disc centre to the inner Lagrangian point. The
eclipse geometry is defined by the inclination $i$ and the mass ratio
$q$.  The mass ratio $q$ defines the shape and the relative size of
the Roche lobes. The inclination $i$ determines the shape and
extension of the shadow of the secondary star as projected onto the
orbital plane. In this paper we adopted the values of Wood \&
Crawford (1986), $q=0.5$ and $i=81\degr$, which correspond to an
eclipse phase width of $\Delta\phi=0.0863$\,cycle. This combination of
parameters ensures that the white dwarf is at the centre of the
map. BHT compared eclipse maps obtained with the above geometry with
maps derived with the geometry of Marsh (1988) ($q=0.58$,
$i=79.5\degr$) and found no perceptible differences in the eclipse
maps.

For the reconstructions we adopted the default of limited azimuthal
smearing of Rutten, van Paradijs \& Tinbergen (1992), which is better
suited for recovering asymmetric structures than the original default
of full azimuthal smearing (cf. Baptista, Steiner \& Horne 1996).  The
reader is referred to BHS and Baptista (2001) for eclipse mapping
simulations with asymmetric sources which show how the presence of
spiral structures affect the shape of the eclipse light curve, and
which evaluate the ability of the eclipse mapping method to
reconstruct asymmetric structures in eclipse maps. The performance of
the eclipse mapping technique for the case of light curves with the
incomplete phase coverage of our data sets is discussed in the
Appendix A.

The statistical uncertainties in the eclipse maps were estimated with
a Monte Carlo procedure (e.g. Rutten et~al.  1992).  For a given
narrow-band light curve a set of 20 artificial light curves is
generated, in which the data points are independently and randomly
varied according to a Gaussian distribution with standard deviation
equals to the uncertainty at that point. The light curves are fitted
with the eclipse mapping algorithm to produce a set of randomized
eclipse maps.  These are combined to produce an average map and a map
of the residuals with respect to the average, which yields the
statistical uncertainty at each pixel.  The uncertainties obtained
with this procedure are used to estimate the errors in the derived
radial intensity and temperature distributions as well as in the
spatially-resolved spectra. 

We shall remark that this procedure only accounts for the statistical
uncertainty in the eclipse maps. Because each outburst stage is
represented by a single light curve, it was not possible to reduce the
influence of flickering in the eclipse shape by the usual averaging of
many individual light curves. Therefore, the eclipse shapes may be
affected by flickering. This could introduce artifacts in the eclipse
maps, which are not accounted for by the above statistical
analysis. Nevertheless, an inspection of the light curves of
Fig.\,\ref{curvas} shows that the out-of-eclipse flickering is of
relatively low amplitude ($\sim$\,10 per cent) and is further reduced
during the eclipse. Moreover, the time scale of individual flares is
much shorter than the time scale of the changes in the eclipse
shape. Since the eclipse mapping modeling is defined by the overall
eclipse shape (see Figs.\,\ref{mapas_cont} and \ref{mapas_lines}) one
should not expect the fast, low amplitude flickering in the light
curve to significantly affect the results.

Light curves and respective eclipse maps at selected passbands for the
continuum and for the strongest emission lines are shown in
Figs.\,\ref{mapas_cont} and \ref{mapas_lines}.

%______________________________________________________________

\section{Results}

\subsection{Disc structure}

Maps of the disc surface brightness, calculated by the maximum entropy
eclipse mapping method, allow us to obtain spatially resolved
information about the disc emission, as well as to probe the spectral
and temporal evolution of the accretion disc. In this section we
discuss the structures in the eclipse maps of the strongest lines in
the spectrum, as well as in selected continuum passbands, and we
compare our results with those of previous studies.

Previous optical studies using eclipse mapping methods (BHS; BHT)
showed evidences for the presence of spiral arms, which are presumably
the consequence of shocks caused by tidal effect of the secondary star
in the extended accretion disc during outbursts. These asymmetric arcs
were seen as structures $\sim 90\degr$ in azimuth extending from the
intermediate to the outer disc regions [$R=(0.2-0.6)R_{L1}$]. BHT
detected spiral arms during the 1993 May outburst, at the day before
the start of our observations. Because our data frame the very last
stages of this outburst, they open the perspective to check how long
after the start of an outburst are the spiral structures still present
in the disc of IP~Peg.

\begin{figure}
  \centering
 \includegraphics[bb=2cm 0cm 21cm 25cm,scale=.65]{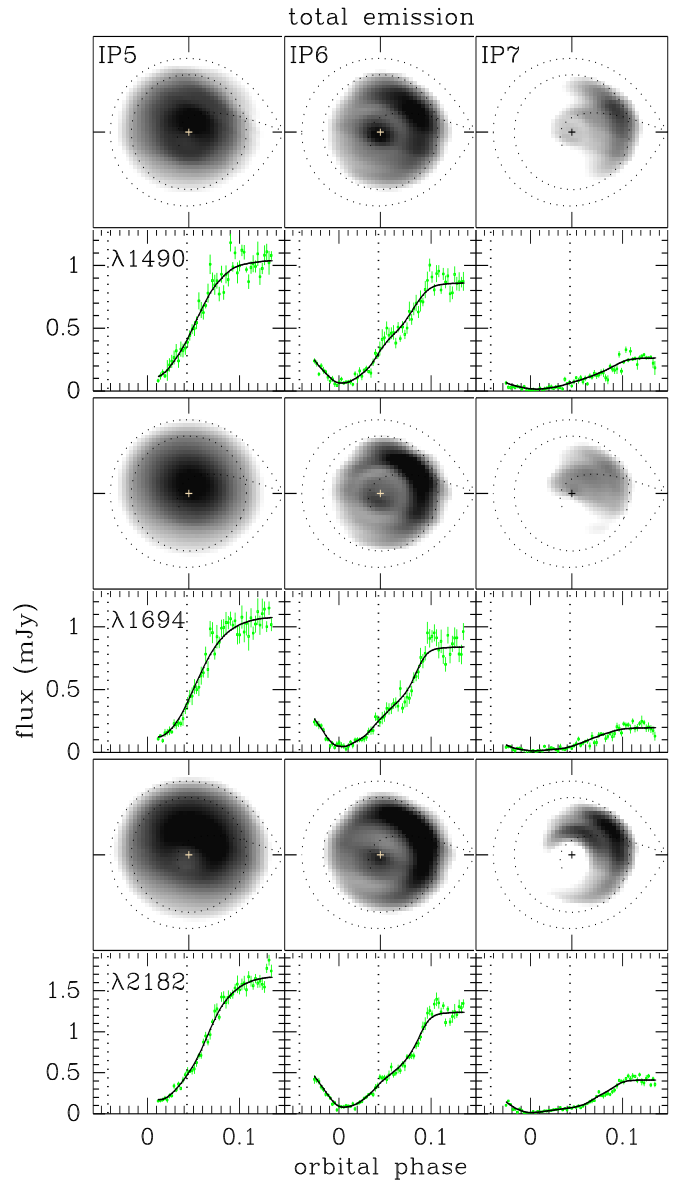}
     \caption{Data and model light curves and eclipse maps for three
     selected continuum passbands. The upper panels of each slide show
     the eclipse maps corresponding to the light curves of the lower
     panels for the three runs in a logarithmic greyscale.  Brighter
     regions are indicated in black; fainter regions in white.  A
     cross marks the center of the disk; dotted lines show the Roche
     lobe, the gas stream trajectory, and a disc of radius
     $0.6\;R_{L1}$; the secondary is to the right of each map and the
     stars rotate counter-clockwise. The lower panels of each slide
     show the data (dots with error bars) and model (solid lines)
     light curves. Vertical dotted lines mark the ingress/egress
     phases of the white dwarf.}
        \label{mapas_cont}
  \end{figure}

Fig.\,\ref{mapas_cont} shows the light curves and respective eclipse
maps for three selected continuum passbands. The IP6 e IP7 light
curves show asymmetries at egress phases, indicating the presence of
more than one emission source or significant intensity variations
along the disc radius. The corresponding eclipse maps show asymmetric
structures in the side of the disc that is moving away from the
secondary star (the upper hemisphere of the eclipse maps in
Fig.\,\ref{mapas_cont}) which can be associated to the gas stream
emission. If this assumptions is right, the fact that the emission
traces the ballistic stream trajectory beyond the bright spot position
is an evidence of gas stream overflow in IP~Peg at this late outburst
stage. The azimuthal extension of the asymmetric structure at the disc
outer regions in IP6 e IP7 shows that part of the energy in the
collision of the stream with the disc edge is dissipated along the
disc rim, in the direction of the gas rotation. It is interesting to
note that while the outburst reaches its final stages, the continuum
maps are progressively dominated by emission along the gas stream
region.

The maximum derivative of the IP5 continuum light curves is displaced
towards later phases with respect to white dwarf egress, indicating
that the maximum of the brightness distribution does not coincide with
the disc centre. Accordingly, the IP5 continuum maps show an
arc-shaped asymmetric structure in the disc side moving away from the
secondary star with azimuthal width $\sim 90\degr$ and spanning a
range of radii $R=(0.2-0.4)R_{L1}$.  The morphology, location,
azimuthal and radial width of this arc-shaped structure are similar to
those of the ``red'' spiral arm seen in BHS. The reader is referred to
the Appendix for simulations with and for a discussion of eclipse
mapping reconstructions of spiral structures from incomplete phase
light curves.

Our eclipse maps do not show evidence of the other (``blue'') spiral
arm. In analogy with the results of BHS, the blue spiral arm should be
located in the region of the eclipse map that is not sampled by the
shadow of the secondary star because of the incomplete phase coverage
of the IP5 data set (see Fig.\,\ref{arcos} in Appendix
A). Since it is not possible to recover the brightness distribution
of disc regions for which there is no information in the shape of the
light curve (e.g. Baptista 2001), it is not surprizing that there is
no evidence of the ``blue'' spiral arm in these maps.

Morales-Rueda, Marsh \& Billington (2000) and Steeghs et~al. (1997)
found evidence of spiral structures in Doppler tomograms of IP~Peg,
respectively 5-6 and 7 days after the onset of an outburst.  Spiral
structures were also clearly seen in the eclipse maps of BHT, obtained
8 days after the onset of this same outburst.  Our IP5 continuum maps
extend the previous results, showing evidence that one spiral arm was
still present in the accretion disc of IP~Peg 9 days after the start
of the outburst.

  \begin{figure}
  \centering
  \includegraphics[bb=2cm 0cm 21cm 25cm,scale=.65]{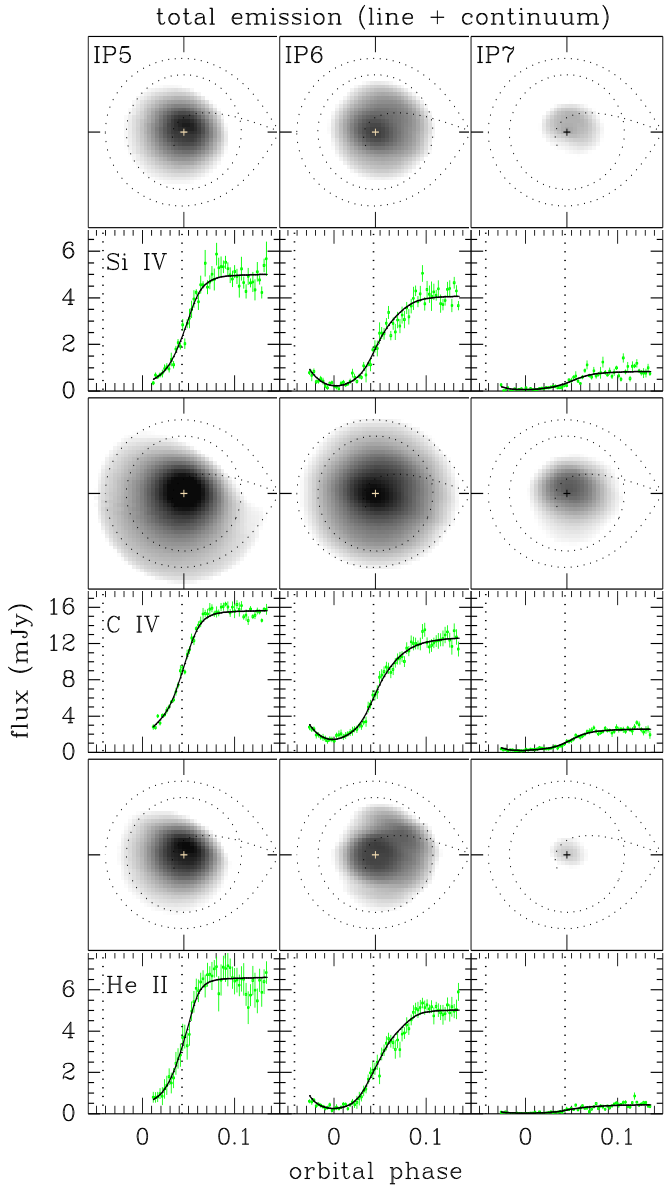}
     \caption{The upper panels of the each slide show the eclipse maps
             for the passband centered in the Si\,IV, C\,IV and He\,II
             emission lines ($v=0\,km/s,
             \Delta{v}=\pm\,1000\,km/s$). The lower panels show the
             data (dots with error bars) and corresponding model
             (solid lines) light curves. The notation is similar to
             that of Fig.\, \ref{mapas_cont}.}
        \label{mapas_lines}
  \end{figure}

Fig.\, \ref{mapas_lines} shows the light curves and respective eclipse
maps for the line centre of the three strongest emission lines,
Si\,IV, C\,IV and He\,II. The velocity-resolved passbands have widths
of $\Delta{v}=2000\,km/s$~(from $-1000$ to $+1000\,km/s$, centered in
$v=0\,km/s)$. The phase of maximum derivative in the light curves for
all runs coincides with the white dwarf egress phase, indicating that
the maxima of the distributions coincide with the disc centre. The
light curves show deep eclipses with smooth egresses.  As a
consequence, the resulting maps show broad brightness distributions
centered at the white dwarf position with no asymmetric structure that
could signal the presence of spiral arms.  This is in contrast with
the results of BHS, the He\,I line maps of which show the same
asymmetric structures observed in the continuum maps.  The strong UV
lines are probably produced in a vertically-extended, opaque
chromosphere + disc wind, which may veil the underlying emission from
the accretion disc and spiral arms.

The eclipse egress occurs perceptibly later in IP6 than in IP5. The
small asymmetry observed in the eclipse egress in the IP6 light curves
produces the emission in the bright spot region in the corresponding
eclipse maps. We note that the line maps contain the line emission
plus any possible underlying continuum. Thus, the presence of
asymmetric structures in the continuum maps and its absence in the
line maps, in addition to the observed spectrum of the uneclipsed
component (see Section 4.4), is a indication of the vertical extension
and large optical depth of the gas from which the lines originate. Our
interpretation is that this opaque and vertically extended gas hides
the structures observed in the continuum, leading to smooth line maps
in which the emission along the bright spot region is much less
visible than in the continuum.

The time sequence of eclipse maps show the progressive fading of both
line and continuum emission as the outburst ends. The intensity of the
maps decrease significantly from run IP6 to IP7. This is particularly
true for the Si\,IV and He\,II line maps, for which the line emission
becomes restricted to the central parts of the disc. In contrast with
the smooth, symmetric line maps, the continuum emission at the end of
the outburst arises mainly at the bright spot position and along the
ballistic stream trajectory.

\subsection{Spatially resolved spectra}

The set of monochromatic eclipse maps allow us to obtain spatially
resolved spectra for the accretion disc. In order to separate the disc
spectrum at different distances from the disc centre, we divided the
disc in concentric annular sections of width $\Delta{R}=0.1R_{L1}$.

\begin{figure}
  \centering
  \includegraphics[bb=1cm 1cm 20cm 21cm,angle=-90,scale=.3]{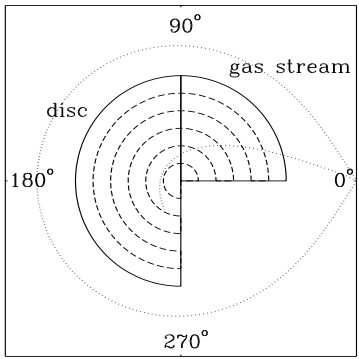}
     \caption{Diagram showing the regions defined as ``disc'' and
  ``gas stream''. The dashed lines mark the annular regions of width
  $0.1\,R_{L1}$ used to extract spatially resolved spectra. Dotted
  lines show the projection of the primary Roche lobe onto the orbital
  plane and the gas stream trajectory. Azimuths are measured with
  respect to the line joining both stars and increase
  counter-clockwise. Four reference azimuths are labeled in the
  figure.  }
        \label{mapa_esq}
  \end{figure}

Motivated by the distinct emission observed in the bright spot and gas
stream region, we separated the eclipse map in two distinct azimuthal
regions: ``disc'' and ``gas stream''. We define the ``gas stream'' as
the map section between $0\degr$ and $90\degr$, and the ``disc'' as
the region between azimuths $90\degr$ and $270\degr$ (see Fig.\,
\ref{mapa_esq}). We do not include the section between $270\degr$ and
$360\degr$ in the ``disc'' region because there is evidence of
contamination by emission from the bright spot in this region (because
of the intrinsic azimuthal smearing effect of the eclipse mapping
technique) and also because there is evidence that the disc has a
non-negligible opening angle (see Section 4.5) and this could
introduce artifacts in the computation of the average flux of each
annulus.

Each point in the spectrum is obtained by averaging the intensity of
all pixels inside the corresponding annulus. The statistical
uncertainties affecting the average intensities are estimated with the
Monte Carlo procedure described in Section 3.2.

  \begin{figure*}
  \centering
  \includegraphics[bb=1cm 6cm 20cm 21cm,angle=-90,scale=.6]{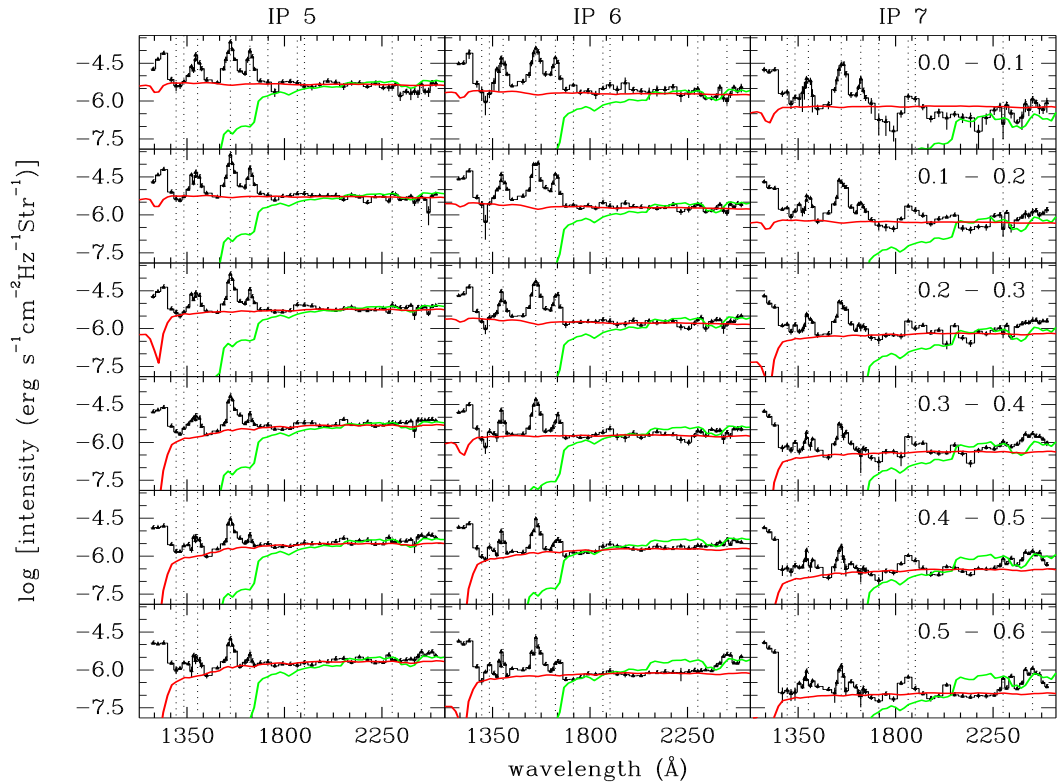}
     \caption{Spatially resolved disc spectra $(90\degr-270\degr)$ for
     a set of six annular regions (indicated in units of
     $R_{L1}$). The best-fit stellar atmosphere model is shown as a
     solid line in each panel (dark gray for the color temperature
     model and light gray for the brightness temperature
     model). Dotted lines mark the major transitions of the spectra.}
        \label{spec_disc}
  \end{figure*}

  \begin{figure*}
  \centering
  \includegraphics[bb=1cm 6cm 20cm 21cm,angle=-90,scale=.6]{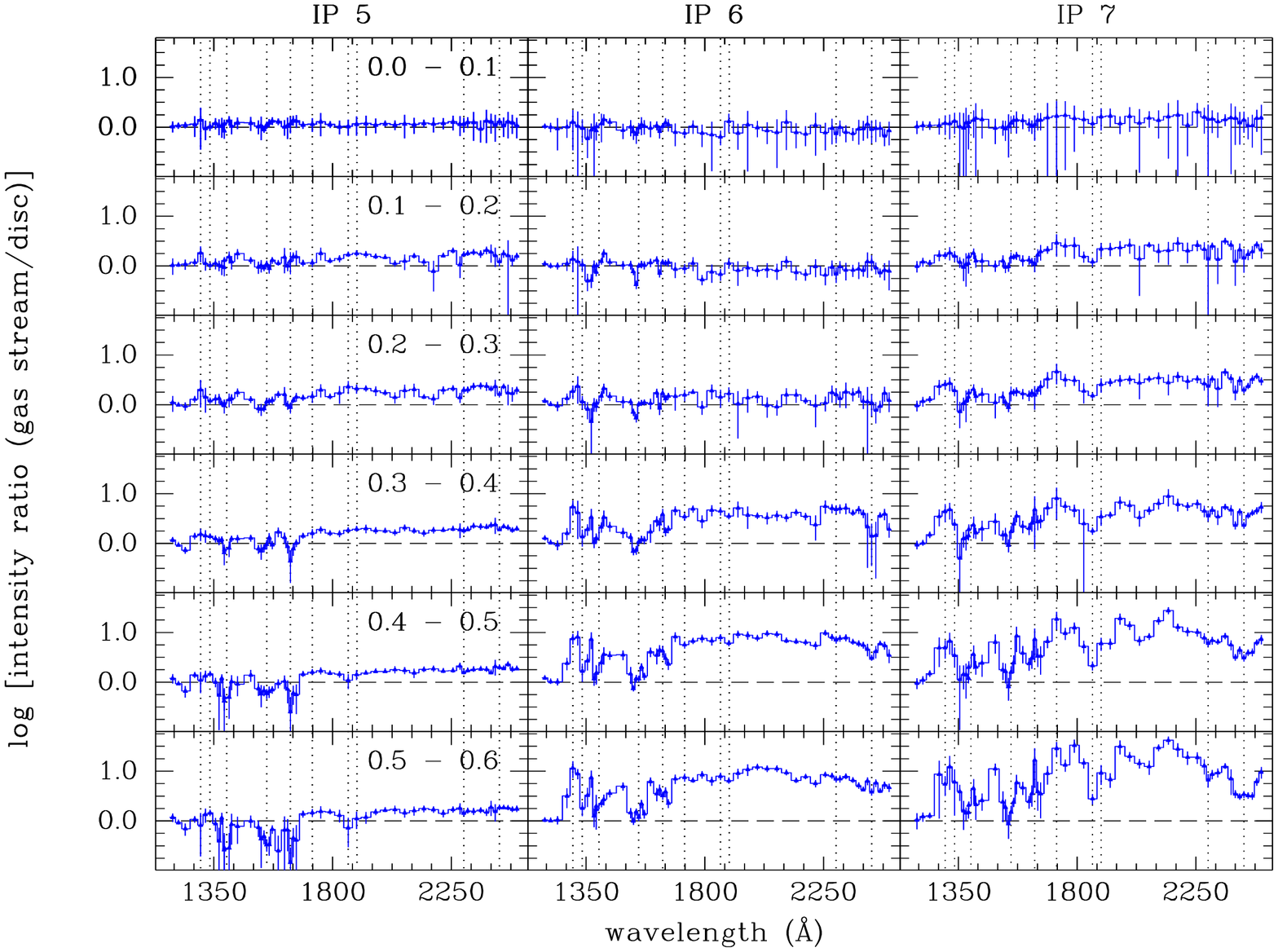}
     \caption{The ratio between the gas stream $(0\degr-90\degr)$ and
     disc $(90\degr-270\degr)$ spectra as a function of radius. The
     notation is similar to that of Fig. \,\ref{spec_disc}.}
        \label{spec_spot}
  \end{figure*}

The spatially resolved disc spectra for the three runs are shown in
Fig.\, \ref{spec_disc}. The spectra display strong lines of Si\,IV,
C\,IV and He\,II that appear in emission at all disc radii. This
behavior is similar to that observed for the H$\alpha$ and He\,I lines
in the contemporary optical spectral mapping experiment of BHS. We
note again the radial extension of the C\,IV emission: in the disc
inner regions the Si\,IV, C\,IV and He\,II lines have comparable
strengths but the Si\,IV and He\,II lines decrease quickly in
intensity with increasing disc radius while the intensity of the C\,IV
line remains approximately at the same level of the inner disc
regions. The He\,II emission is concentrated in the disc centre and is
possibly associated to photoionization by radiation from the boundary
layer (Marsh \& Horne 1990).

Patterson \& Raymond (1985) discussed how reprocessing of the soft
X-rays generated in the boundary layer can lead to He\,II emission
when the accretion rate exceeds $\dot{M} \gtrsim
1.6\times10^{-9}\,M_{\sun}\,yr^{-1}$. This value is consistent with
our results for IP~Peg in the final stages of the outburst (see
Section 4.5). Webb et al. (1999) derived an accretion rate of
$8.0\times10^{-9}\,M_{\sun}\,yr^{-1}$ at outburst maximum, while Marsh
(1988) estimated a rate of $2.2\times10^{-10}\,M_{\sun}\,yr^{-1}$ in
quiescence. The He\,II emission practically disappears in IP7, with
only a residual emission from the innermost disc regions
remaining. It is possible that the mass accretion rate in the inner
disc has reduced to a level insufficient for photoionization.

Ly$\alpha$ also appears in emission at all disc radii. However, the
light curve for this line is largely affected by geo-coronal
emission and the results must be looked at with caution and
skepticism.

For the three runs, the slope of the continuum varies in a consistent
way with the distance to the disc centre. The spectra are bluer in the
central parts and redder in the outer parts, indicating the existence
of a radial temperature gradient in the disc.

In order to investigate the emission along the gas stream trajectory,
we calculate the ratio between the spectrum of the gas stream and of
the disc at same radius (Fig.\,\ref{spec_spot}). The results reveal
that the spectrum of the gas stream is indistinguishable from the disc
spectrum in the inner disc regions, but becomes clearly distinct from
the disc spectrum in the intermediate and mainly in the outer disc
regions. For a large range of radii ($R\gtrsim0.3\,R_{L1}$) the
continuum in the gas stream region is stronger than in the disc, in
particular in IP7. In contrast to the observed in the disc spectra,
where the lines appear in emission at any radius, in the ratio of the
gas stream to the disc spectra the Si\,IV, C\,IV and He\,II lines
appear in absorption in the regions where there is distinct gas stream
emission in the three runs.

As discussed in Section 4.1, the distinct emission along the gas
stream trajectory is an evidence of gas stream overflow beyond the
point of impact with the disc edge.

The fact that the lines appear in absorption in the stream spectra
ratio suggests the presence of matter outside of the orbital plane
(e.g. chromosphere + wind) or that the disc has a non-negligible
opening angle (flared disc). Horne et al. (1994) observed strong C\,IV
emission out of the orbital plane in the dwarf nova OY~Car in
quiescence and suggested that this emission may arise from magnetic
activity on the surface of the quiescent disc (Horne \& Saar 1991), or
from resonant scattering in a vertically-extended region well above
the disc. There is strong evidence of the existence of disc winds in
systems with high mass accretion rates, as is observed in IP~Peg in
outburst (see Section 4.4)[Horne et al. 1994]. The disc wind
hypothesis was used by BHS to explain the behavior of the uneclipsed
component (Section 4.4), while the alternative hypothesis of a flared
disc is suggest from the radial run of the solid angles from the fits
of model spectra (Section 4.5).

\subsection{The emission lines}

\begin{figure*}
  \centering
  \includegraphics[bb=1cm 7cm 20cm 21cm,angle=-90,scale=.5]{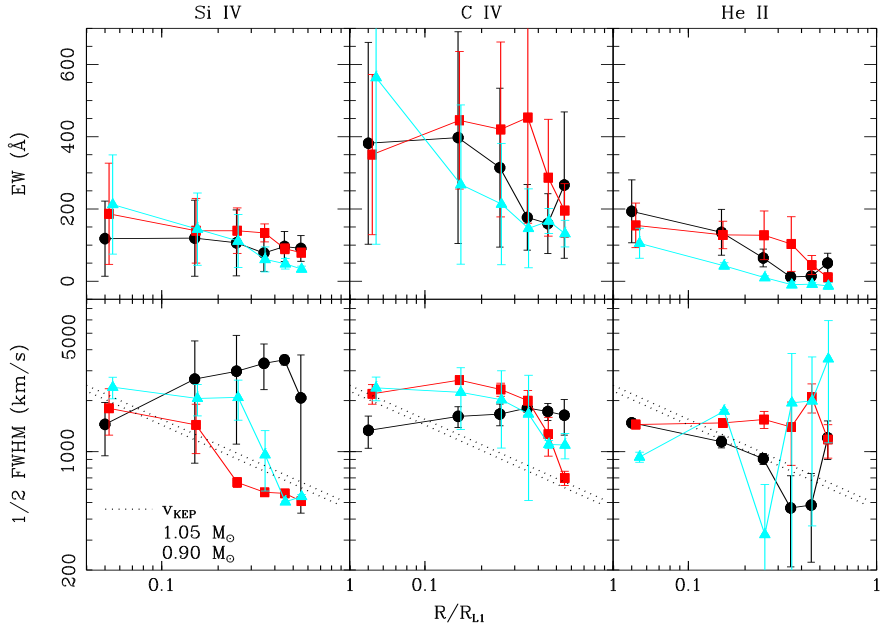}
     \caption{Radial run of the equivalent width (EW) and full width
       half maximum (FWHM) of the most prominent emission lines as a
       function of time for the three runs. The run IP5 is shown as
       black lines with filled circles, IP6 as dark gray lines with
       filled squares, and IP7 as light gray lines with filled
       triangles. The dotted lines in the lower panels show the
       expected behavior for gas in Keplerian orbits around a primary
       star of mass $1.05\,M_{\sun}$ (upper line) and $0.90\,M_{\sun}$
       (lower line).}
        \label{ew}
  \end{figure*}

Fig.\,\ref{ew} shows the radial run of the equivalent width (EW) and
full-width-half-maximum (FWHM) of the most prominent emission lines as
a function of time.  The lines are stronger in the inner disc regions
and decrease in strength with increasing disc radius. C\,IV is the
dominant line with an EW $\simeq 400$~\AA\ at 0.1 $R_{L1}$, while
Si\,IV and He\,II, have EW $\simeq 150$~\AA\ at the same radius.

The He\,II emission is more concentrated in the inner regions and
practically disappears in the continuum in the outer disc
regions. This behavior is more pronounced in IP7. This underscores the
conclusions drawn in Sections 4.1 and 4.2 that in the last stages of
the outburst the He\,II emission arises mostly from the disc
centre. The last three points of the FWHM diagram of He\,II are very
noisy and not reliable because the He\,II line is hardly discernible
in the outer disc regions. In contrast, the Si\,IV and C\,IV emission
is spread over the disc.

The FWHM of the C\,IV line is approximately constant with radius for
the three runs and the derived values (1/2 FWHM $\simeq2000\,km/s$)
are comparable to those found for others dwarf novae (e.g. OY\,Car,
Horne et al. 1994). For Si\,IV the FWHM is approximately constant with
radius in IP5, while in He\,II the FWHM is essentially constant with
radius in IP6. In the other runs (IP5 and IP7) the FWHM of He\,II has
very large error bars in the outer disc regions and the results are
unreliable.

In IP5 we measure velocities of $\simeq 2000\,km/s$ for C\,IV and
Si\,IV in the outer disc regions. As a general trend, all lines show
high FWHM values at large radii and a flat FWHM radial distribution
that is markedly different from the expected behavior of gas in
Keplerian orbits, where $v \propto R^{-1/2}$. In particular, the
inferred FWHM values of the outer disc regions are factor of $3-4$
times larger than the Keplerian expectation.  The high values of the
FWHM measured in the outer disc regions indicate that the lines are
not being emitted by gas in Keplerian orbits, but possibly in a
chromosphere plus disc wind. The wind hypothesis was evoked in
Sections 4.1 and 4.2 to explain the absence of structures related to
spiral arms and the distinct emission of the gas stream region in the
line maps.

\subsection{The uneclipsed component}

The uneclipsed component was introduced in the eclipse mapping method
to account for the fraction of the total light which is not coming
from the accretion disc plane (e.g. light from the secondary star or
from a vertically-extended disc wind)~[Rutten et al. 1992].  

  \begin{figure}
  \centering
  \includegraphics[bb=3cm 11cm 18cm 16cm,angle=-90,scale=.4]{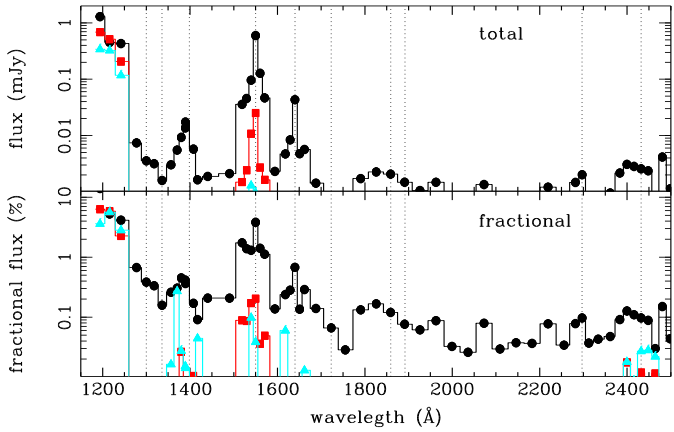}
     \caption{Spectrum of the uneclipsed component. Top panel:
     total contribution. Bottom panel: fractional contribution. The
     notation is similar to that of Fig. \ref{ew}.}
        \label{neclip}
  \end{figure}

We estimate the fractional contribution of the uneclipsed component to
the total flux dividing the flux of the uneclipsed light by the
average out of eclipse level at the corresponding
passband. Fig.\,\ref{neclip} shows the spectrum of the uneclipsed
component as well as its fractional contribution as a function of
wavelength. It is dominated by strong C\,IV and $Ly\alpha$ emission
lines (however, most of the Ly$\alpha$ emission is probably of
geo-coronal origin).

Previous studies in the infrared indicated that the secondary of
IP~Peg is a cool star of spectral-type M5 with a temperature T
$\simeq4000$ K (Catal\'an, Smith \& Jones 2000). Therefore, one does
not expect this star to contribute to the emission in the ultraviolet
range.  Possible irradiation effects during outburst should not affect
the contribution of the secondary star to the uneclipsed light because
during eclipse one sees essentially light from the side opposed to the
white dwarf, which should be free from illumination by the disc and
boundary layer. Moreover the facts that the lines appear in emission
and that the uneclipsed component decreases toward the end of the
outburst indicate that this emission is related to the outburst,
discarding the hypothesis of a contribution from the secondary star.

BHT showed that the uneclipsed optical continuum contributes an
increasing fraction of the total flux for longer wavelengths, reaching
10 per cent at the red end of the spectrum ($\sim6700$\,\AA). This
suggest that the Paschen jump is in emission in the uneclipsed
component. Unfortunately, our spectral coverage does not allow to test
if the Balmer jump is also in emission. 

The uneclipsed component of C\,IV accounts for 4 per cent of the total
light in IP5 but decreases below the 0.5 per cent level in the other
runs (Fig.\,\ref{neclip}), signaling that the vertical extension of
the C\,IV line emitting region decreases rapidly as the outburst
approaches its end.  The other lines and the continuum in the
uneclipsed spectrum yield negligible fractional contributions
($\lesssim$ 1\,per cent) to the total light at the corresponding
wavelengths. For IP5, the continuum contributes $\simeq0.1$ per cent
of the total light and decreases to still smaller values at later
stages ($\lesssim0.01$ per cent). The intensity of the uneclipsed
component decreases much faster than that of the disc emission. If the
uneclipsed component arises from a vertically-extended disc
chromosphere + wind, the faster reduction of its strength in
comparison with the disc emission can be interpreted as caused by the
reduction of the vertical extension of this region. This corroborates
the idea that the wind is associated to and powered by the outburst
itself.

    \begin{figure}
    \centering
    \includegraphics[bb=1cm 6cm 20cm 21cm,angle=-90,scale=.6]{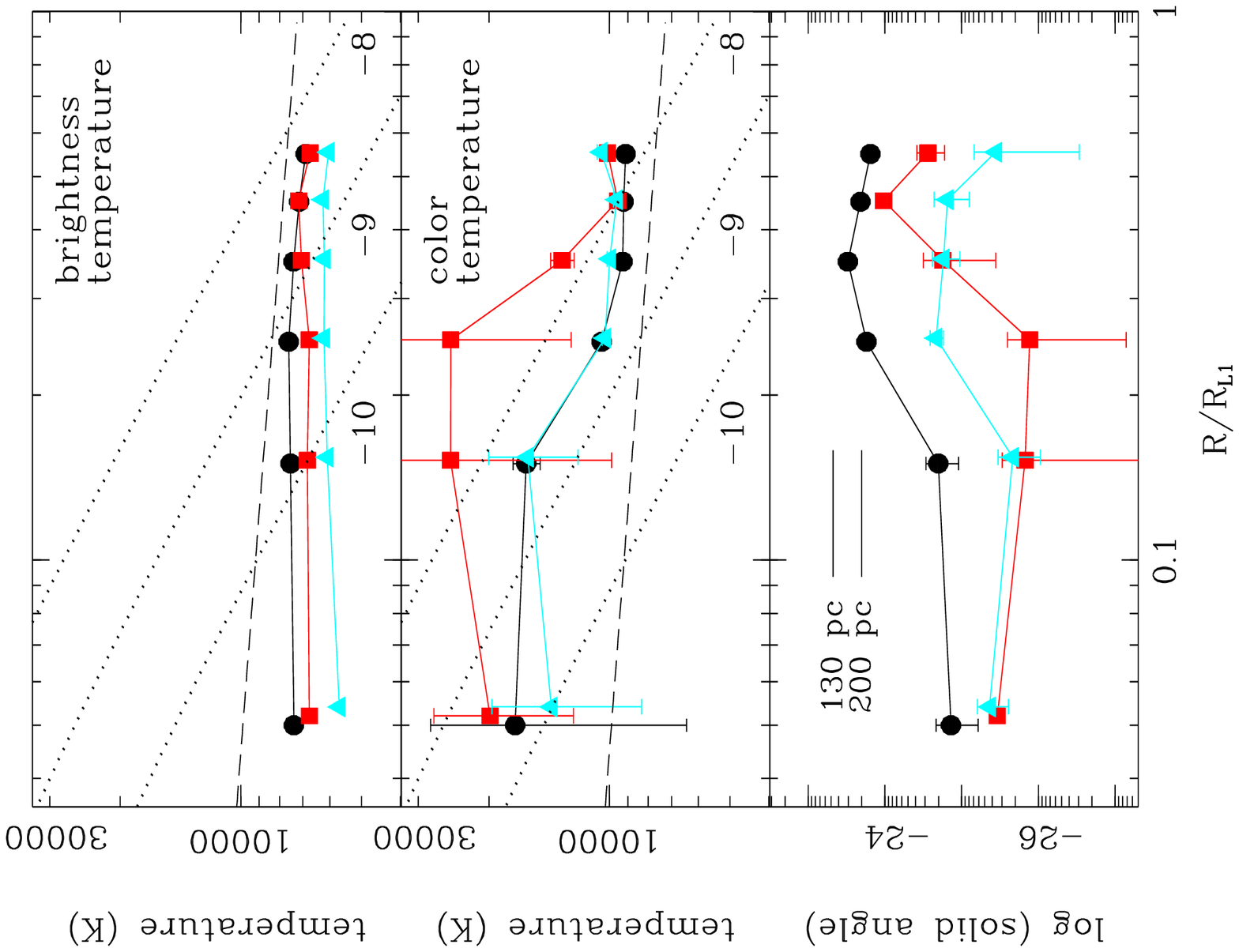}
       \caption{Top panel: the radial run of the brightness
  temperature of the best-fit stellar atmosphere models to the disc
  spectra as a function of time for the three runs. Dotted lines show
  steady-state disc models for mass accretion rates of $\rm \dot{M}$ =
  $10^{-8}$, $10^{-9}$ and $10^{-10}$ M$_{\sun}{\rm \: yr^{-1}}$.  The
  dashed line show the critical temperature above which the gas should
  remain in a steady, high-\.{M} regime. Middle panel: The radial run
  of the color temperature, the notation is the same to that of top
  panel. Bottom panel: the solid angle of the best-fit color
  temperature model in each case.  Horizontal ticks mark the expected
  solid angle for $\rm i=81\degr$, $\rm R_{L1}=0.81$ R$_{\sun}$ and
  distances of 130\,pc and 200\,pc. The notation is similar to that of
  Fig. \ref{ew}.}
          \label{temp}
    \end{figure}

\subsection{The radial run of the temperature and solid angle}

The fit of model spectra to the spatially resolved continuum disc
spectra allows us to infer the radial run of temperature and solid
angle. It is then possible to compare the derived radial temperature
distribution with the $T\propto{R^{-3/4}}$ law expected for
steady-state optically thick discs, as well as to follow the evolution
of this distribution in the late stages of the outburst.

We fitted blackbodies, simple LTE H\,I and stellar atmosphere models
with metalicities log[M]/[H]=0.0, 0.5 and 1.0 (Kurucz 1979) to the
observed spatially resolved spectra. The fits do not take into
consideration possible effects of limb darkening and do not include
the spectral regions containing emission lines. The statistical
uncertainties in the values of the temperature and solid angle
extracted from the spectral fits were estimated with a Monte Carlo
procedure (Section 3.2).

The LTE H\,I fitted models resulted in high values for the column
density, of order of $10^{21}\rm\,particles \,cm^{-2}$, indicating that
the disc gas is optically thick. The derived values of the temperature
and of the solid angle are indistinguishable from those derived
assuming blackbody emission.

The best-fit for blackbodies yield fits with
$\chi^{2}\simeq4-5$. These fits lead to high temperatures, of the
order of T $\simeq60000$ K in the inner disc regions and T
$\simeq20000$ K in the outer regions. The steady-state optically thick
discs models corresponding to these temperatures imply in mass
transfer rates ($\dot{M}\gtrsim10^{-7}\,M_{\sun}\,yr^{-1}$) well above
those expected for dwarf novae in outburst and even high $\dot{M}$
nova-like systems (Warner 1995).

The fits of stellar atmosphere models were performed in two different
ways. First, we adopted a distance of 130 pc to IP~Peg (Martin, Smith
\& Jones 1987). We fixed the solid angle at the corresponding value
for all disc annuli and fitted the model spectra to find the
brightness temperature of best-fit -- i.e., the temperature of the
model spectrum that produces the same intensity of the observed
spectrum. The results are shown as solid light gray lines in
Fig.\,\ref{spec_disc}. These models yield systematically poor fits to
the blue side of the spectra, largely under estimating the intensities
shortward of $\simeq1800$ \AA, and result in $\chi^{2}\gtrsim100$. The
radial run of the brightness temperatures obtained from these fits are
shown in the upper panel of Fig.\,\ref{temp}. Our inferred brightness
temperatures decrease monotonically from $\simeq 8000$ K in IP5 to
$\simeq 6000$ K in IP7. The temperatures and the slope of the
distribution are similar to those obtained by Bobinger et al. (1997)
and by BHT, who found flat radial temperature distributions for IP~Peg
on the decline from outburst, with brightness temperatures
respectively of $5000-9000$ K and of $\simeq5000$ K.

The inferred disc brightness temperatures are systematically higher
than those of BHT although our data correspond to a later outburst
stage. Together with the clear mismatch of the brightness temperature
models to the short wavelengths of the observed spectra, this is an
indication that (blackbody) brightness temperatures are a not a good
approximation to the disc effective temperatures in this
case. 

Therefore, we included the solid angle as a free parameter of the
model, and fitted temperature and solid angle independently for each
annulus. This provides a color temperature for each annulus -- we fit
the shape of the spectrum and we let the solid angle account for any
discrepancy in the intensity. The best-fit color temperature stellar
atmosphere models are shown as solid dark gray lines in
Fig.\,\ref{spec_disc}.

The color temperature models provide a much better fit to the data
than the brightness temperature models in all cases. Among the stellar
atmosphere models tested (Kurucz 1979), those of solar metalicity
(log[M]/[H]=0.0) yield fits that are systematically better than those
of other metalicities, resulting in $\chi^{2}$ values comparable to
those obtained with blackbody models. The radial run of the derived
color temperatures and the corresponding solid angles are shown,
respectively, in the middle and lower panels of Fig.\,\ref{temp}. The
disc color temperature distributions are also flatter than the
expected $T\propto{R^{-3/4}}$ law for steady-state optically thick
discs models. However, the color temperatures are systematically
higher than the brightness temperatures, with values of
$T\simeq{20000}$ K in the inner disc regions and $T\simeq{9000}$ K in
the outer regions.

The derived color temperatures at the end of the outburst (IP7) are
everywhere higher than the critical temperature above which the gas
should remain in a steady, high-mass accretion regime (Warner 1995).
If the color temperatures are representative of the disc effective
temperatures, this seems hard to reconcile with the disc instability
model, which predicts that most of the disc gas should be back to the
low viscosity, low temperature branch of the thermal limit cycle at
these late outburst stages (e.g. Lin, Faulkner \& Papaloizou 1985).

The comparatively low color temperatures obtained for the inner disc
regions may be a consequence of a geometrically thick (flared) disc
during outburst. If in IP~Peg the inner disc regions are hidden from
view by a thick disc rim, this will result in artificially low
temperatures for the inner (hidden) disc regions. At the high
inclination angle of IP~Peg $(i=81\degr)$ a moderate disc openning
angle, $\alpha\simeq10\degr$, would suffice to produce this
effect. Webb et al. (1999) and Morales-Rueda (2000) argued that during
outbursts the disc of IP~Peg is flared with an opening angle
$\alpha\lesssim15\degr$.

The solid angles corresponding to the color temperature stellar
atmosphere models are shown in the lower panel of
Fig.\,\ref{temp}. The solid angle increases with radius suggesting
that the outbursting disc of IP~Peg is possibly flared. Furthermore,
the time evolution of the radial run of the solid angles suggests that
the disc thickness decreases toward the end of the outburst. However,
these results must be looked at with some reservations because the
inferred solid angles are typically an order of magnitude smaller than
the predicted for the expected distance to IP~Peg ($130-200$ pc).

Because of the larger disc opening angle inferred for IP5, the inner
disc regions may be hidden from view by a thick disc rim. This result
can explain some apparent contradictions, p.ex. the greater inner disc
temperatures found for IP6 in comparison with IP5.

\section{Conclusions}

The main results of our spectroscopic study of IP~Peg during the
decline of the 1993 May outburst can be summarized as follows:

   \begin{enumerate}
\item The IP5 continuum maps show an asymmetric structure at a similar
azimuth and radius as one of the spiral arms previously seen. This
suggests that the spiral arms are still present in the IP~Peg disc 9
days after the onset of the outburst, although the incomplete eclipse
phase coverage of this run prevents a more conclusive
statement. Simulations show that the incomplete phase coverage of our
data does not allow us to reconstruct the other spiral arm. The other
runs show no clear evidence of spiral arms.

\item The pronounced emission along the gas stream trajectory in IP6
and IP7 is an evidence of gas stream overflow.

\item Spatially resolved spectra reveal that Si\,IV, C\,IV and He\,II
are in emission at all disc radii.  The lines are stronger in the
inner disc regions and decrease in strength with increasing disc
radius.  The FWHM of the C\,IV line is approximately constant with
radius, in contrast to the expected $v \propto R^{-1/2}$ law for gas
in Keplerian orbits. Similar behavior is observed in Si\,IV. These
lines are probably emitted in a vertically-extended region
(chromosphere + wind).

\item The uneclipsed component is dominated by strong C\,IV emission,
which contributes about 4 per cent of the line flux in IP5. The
fractional contribution of the uneclipsed component decreases quickly
during the outburst later stages. The interpretation is that the
uneclipsed component originates in a vertically-extended gas, the
vertical extension of which decreases rapidly at the end of the
outburst.

\item Stellar atmosphere models fitted to the spatially resolved
spectra lead to color temperatures in the range $T\simeq{20000}$ K in
the inner disc regions and $T\simeq{9000}$ K in the outer regions.
For the three runs, the radial distribution of the disc color
temperatures is flatter than the expected $T\propto{R^{-3/4}}$ law for
steady-state optically thick discs. The derived color temperatures at
the end of the outburst are everywhere higher than the critical
temperature above which the gas should remain in a steady, high-mass
accretion regime, in contradiction with predictions of the disc
instability model for dwarf nova outburst.

\item The solid angles resulting from the spectral fit increase with
disc radii. The radial run of the solid angles suggests that the disc
is flared, and decreases in thickness toward the end of the outburst.

   \end{enumerate}

\begin{acknowledgements}
%      \emph {}
      In this research we have used, and acknowledge with thanks, data
      from AAVSO International Data base.  This work was partially
      supported by CNPq/Brazil through the research grant 62.0053/01-1
      -- PADCT III/Milenio. RB acknowledges financial support from
      CNPq/Brazil through grant n. 300.354/96-7. RKS acknowledges
      financial support from CAPES/Brazil.
\end{acknowledgements}

\appendix

\section{Reconstruction of maps from light curves with incomplete phase coverage}

Here we address the reliability of eclipse mapping reconstructions of
asymmetric brightness distributions such as spiral arms for the case
of light curves with incomplete phase coverage.

\begin{figure*}
   \centering \includegraphics[bb=-2cm 0cm 19cm 28cm,scale=.85]{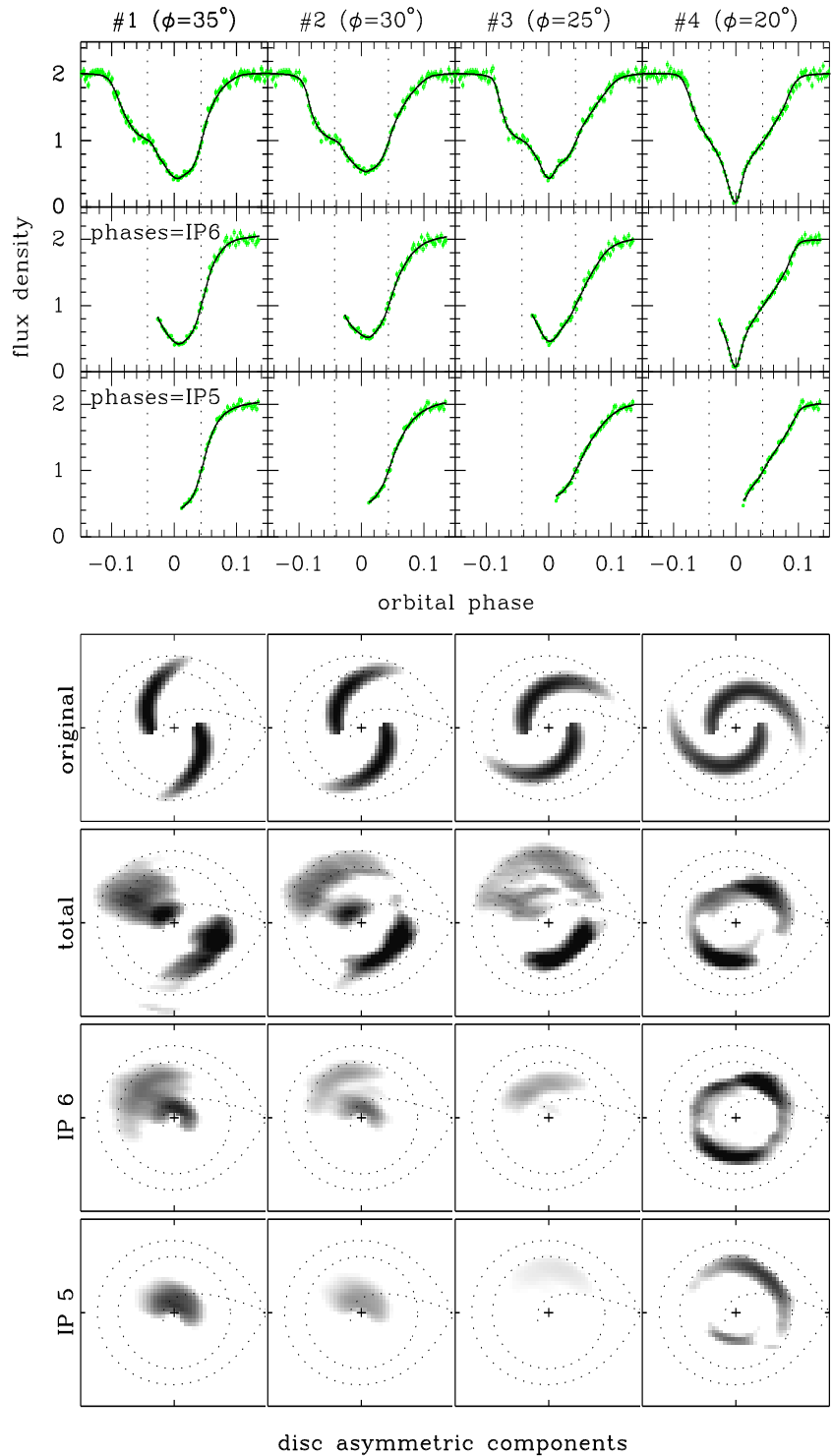}
      \caption{Top panels: The synthetic light curves generated with
$S/N=35$, full phase coverage and with the phase set of runs IP6 and
IP5. Vertical dotted lines mark the ingress/egress times of the white
dwarf. Bottom panels: The configurations of the spiral arms for four
different opening angles used in the reconstructions and the eclipse
maps obtained from the curves of the top panels. A cross marks the
center of the disk; dotted lines show the Roche lobe, the gas stream
trajectory, and a disc of radius $0.6\;R_{L1}$ (for the parameters of
IP~Peg); the secondary is to the right of each map and the stars
rotate counterclockwise.}
         \label{simul}
   \end{figure*}

First, we created eclipse maps of $51\times51$ pixels with two
asymmetric structures at a similar position to the spiral arms
observed in IP~Peg by BHS. It is expected that the spiral arms become
more tighty wound as the disc material cools along the decline from
outburst maximum (e.g. Sawada, Matsuda \& Hashisu 1986; Sato, Sawada
\& Ohnishi 2003). Because our observations of IP~Peg covered the late
stages of the outburst, we used configurations with four different
opening angles $\phi$ for the spiral arms (e.g. Harlaftis et
al. 2004), to cover a range of possible stages of spirals winding
($\phi=35, 30, 25$ and $20\degr$). The artificial spiral models are
shown in Fig.\,\ref{simul}.

We adopt the geometry of Wood \& Crawford (1986) for IP~Peg
($i=81\degr$ and $q=0.5$) to simulate eclipses
of these brightness distributions and to generate synthetic light
curves. We added Gaussian noise to the synthetic light curves to
simulate real data with a signal-to-noise ratio $(S/N)$ comparable to
that of the IP~Peg data ($S/N=35$).

Since the light curves of IP~Peg have incomplete phase coverage, we
truncated the synthetic light curves to produce the set of orbital
phases of runs IP5 and IP6 (the phase coverage of IP7 is identical to
IP6). Fig.\,\ref{simul} shows the light curves for a complete phase
coverage and for the set of phases of IP5 and IP6. The shape of the
eclipse changes significantly with the opening angle of the spiral
arms. We applied eclipse mapping techniques to the three light curves
(one with complete phase coverage, other with the phase set of IP6 and
the last with the phase set of IP5) for each model and compared the
results with the original maps.

\begin{figure}
   \centering \includegraphics[bb=1cm 0cm 19cm 26cm,scale=.45]{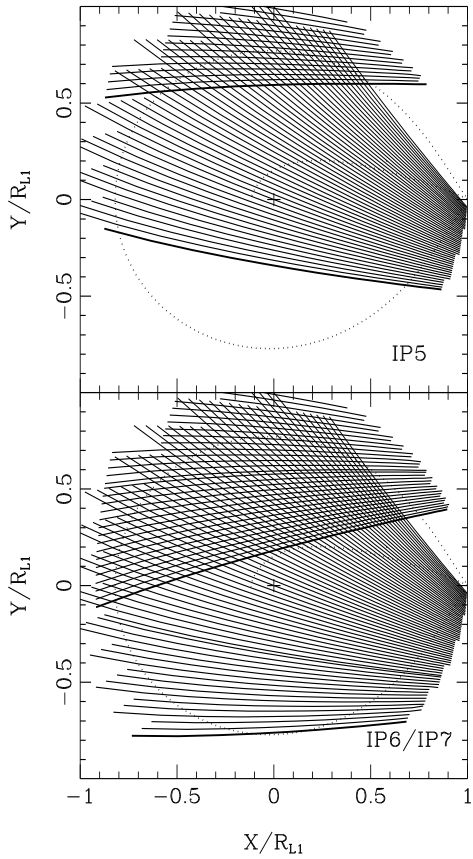}
      \caption{The eclipse geometry of IP Peg for the set of phases of
runs IP5 and IP6/IP7.  Arches (solid lines) connect points of constant
ingress and egress phases.  The pair of thick lines in each panel
shows the ingress/egress arches for the first phase of the
corresponding light curve ($\phi= +0.012$ for IP5 and $\phi= -0.027$
for IP6/IP7).  Dotted lines depict the primary Roche lobe and the
ballistic stream trajectory.}
         \label{arcos}
   \end{figure}

Fig. \ref{arcos} illustrates the eclipse geometry of IP Peg for the
set of phases of runs IP5 and IP6/IP7.  Arches drawn on the eclipse
map connect points of constant ingress and egress phases.  The pair of
thick lines in each panel shows the ingress/egress arches for the
first phase of the corresponding light curve ($\phi= +0.012$ for IP5
and $\phi= -0.027$ for IP6/IP7).  Each pair of arches outlines a
region of the disc that is being occulted by the shadow of the
secondary star at a given phase.  The ability of the eclipse mapping
method to construct a two-dimensional surface brightness map from a
one-dimensional eclipse light curves stems for the fact that
ingress/egress arches intersect with each other at large angles (Horne
1983).

The effect of incomplete phase coverage is to reduce the information
available to reproduce structures in parts of the accretion disc.  For
IP6/IP7, the disc surface is fully sampled by the ingress/egress
arches although about 60 per cent of the disc is only sampled by
egress arches.  On the other hand, about 1/3 of the disc surface is
not sampled by ingress or egress arches with the largely incomplete
phase set of IP5.  There is no information in the eclipse shape about
the brightness distribution of the disc regions not sampled by the
grid of arcs.  Therefore, a spiral arm in the disc side approaching
the secondary star (the lower hemisphere of the maps in
Fig. \ref{arcos}) will not affect the shape of an eclipse light curve
with the phase set of IP5 and will not be recovered by the eclipse
mapping method. 

Fig.\,\ref{simul} shows the original spiral model maps, the artificial
light curves generated (with complete phase coverage and with the
phase sets of IP5 and IP6) and the asymmetric component of the
corresponding reconstructions.  The asymmetric component of each map
is computed separating the radial intensity profile into radial bins
of equal width. We then sort the intensities in each bin and fit a
smooth spline function to the lower quartile of the intensities in
each bin. The fit is then subtracted from each pixel and the resulting
map is saved as the asymmetric component. This process essentially
removes the baseline of the radial profile, leaving all azimuthal
structure in the asymmetric map. We further set to zero all negative
intensities in the asymmetric map.

The asymmetric structures in model \#1 are reasonably well reconstructed
with the full phase coverage light curves (Fig.\,\ref{simul}). As a
consequence of the intrinsic azimuthal smearing of the eclipse mapping
technique (see Baptista 2001 for more details), the arms appear
smeared into a ``butterfly''-shape structure (Harlaftis et~al. 2004). 
However, its position, radial and azimuthal extension are
reasonably well reconstructed. Because the light curve with the set of
phases of IP6 lacks most ingress phases the technique is not able to
properly reconstruct the arm located in the disc region that approaches
the secondary (the ``blue'' arm). The other spiral arm (``red'' arm)
is recovered in a similar fashion to the full coverage light curve.
For the largely incomplete IP5 light curve, the red arm appears
distorted into a crescent-shape asymmetry in the disc side moving away
from the secondary star. As expected, the blue arm is not recovered in
this case.

   \begin{figure}
   \centering \includegraphics[bb=6cm 2cm 9cm 12cm,scale=.69]{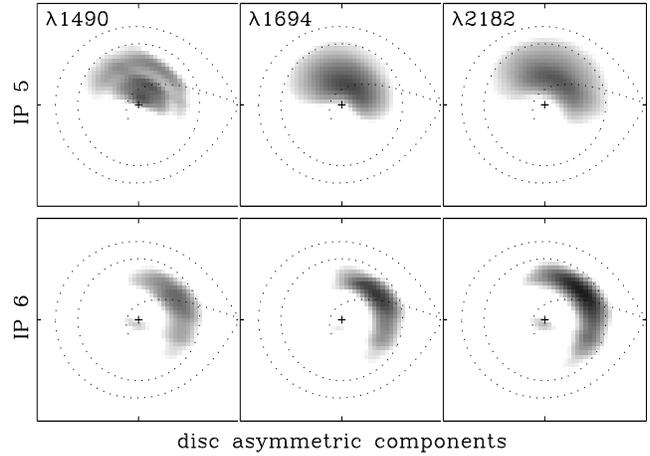}
      \caption{The asymmetric component of the maps for selected
continuum passbands of the runs IP5 and IP6.}
         \label{compara}
   \end{figure}

The reconstructions of model \#2 are similar to those of model \#1
in all three cases. 

As the opening angle of the spirals changes, their orientation and
position in the accretion disc also changes, affecting the ability of
the eclipse mapping method to recover them.
For model \#3, the red arm is well recovered for the IP6 light curve
but is hardly discernible in the extreme case of the IP5 light curve.
For model \#4, both spiral arms are well recovered with the IP6 light
curve, whereas the reconstruction from the IP5 light curve starts to
show hints of the blue arm.

For comparison, in Fig.\,\ref{compara} we show the asymmetric component
of the eclipse maps for selected continuum passbands of the IP5 and 
IP6 runs.

The observed asymmetries in the IP5 continuum maps are very similar to
those seen in the reconstructions of models \#1 and \#2 for the same
set of phases, and can be interpreted as the distorted reproduction of
the ``red'' spiral arm from a light curve with an incomplete set of
phases (missing all the eclipse ingress).

In summary, the simulations show that the eclipse mapping technique
fails to recover structures in disc regions for which there is no
information in the shape of the eclipse. However, asymmetric 
structures such as spiral arms can still be partly detected in eclipse
maps derived from light curves with incomplete phase coverage such as
those of the IP5 run.

\end{document}